\documentclass[11pt,twoside,letterpaper]{article} 
\usepackage{times,fancyhdr}
\usepackage[dvips]{graphicx}
\usepackage{enumitem}


\makeatletter
\def\cleardoublepage{\clearpage\if@twoside \ifodd\c@page\else%
    \hbox{}%
    \thispagestyle{empty}%
    \newpage%
    \if@twocolumn\hbox{}\newpage\fi\fi\fi}
\makeatother

\def\figurename{Figure}
\makeatletter
\renewcommand{\fnum@figure}[1]{\figurename~\thefigure.}
\makeatother

\def\tablename{Table}
\makeatletter
\renewcommand{\fnum@table}[1]{\tablename~\thetable.}
\makeatother

\begin{document}
\title{
{\begin{flushleft}
\vskip 0.45in
{\normalsize\bfseries\textit{Chapter~1}}
\end{flushleft}
\vskip 0.45in \bfseries\scshape Nuclear constraints on gravitational waves from deformed pulsars}}
\author{\bfseries\itshape Plamen G. Krastev\thanks{E-mail address: plamenkrastev@fas.harvard.edu}\\
Harvard University, Faculty of Arts and Sciences, Research Computing\\
38 Oxford Street, Cambridge, MA 02138, U.S.A.\\
\\\\
\bfseries\itshape Bao-An Li\thanks{E-mail address: Bao-An.Li@tamuc.edu}\\
Department of Physics and Astronomy, Texas A\&M University-Commerce,\\
 P.O. Box 3011, Commerce, TX 75429, U.S.A.}
\date{\today}
\maketitle \thispagestyle{empty} \setcounter{page}{1}
\thispagestyle{fancy} \fancyhead{} \fancyhead[L]{Gravitational Waves: Exploration, Insights and Detection,\\
Editor: Nadya Columbus et al., pp. {\thepage-\pageref{lastpage-01}}} 
\fancyhead[R]{ISBN 0000000000  \\
\copyright~2010 Nova Science Publishers, Inc.} \fancyfoot{}
\renewcommand{\headrulewidth}{0pt}

\vspace{2in}


\newpage

\pagestyle{fancy} \fancyhead{} \fancyhead[EC]{Plamen G. Krastev
and Bao-An Li} \fancyhead[EL,OR]{\thepage} \fancyhead[OC]{Nuclear constraints on gravitational waves from deformed pulsars}
\fancyfoot{}
\renewcommand\headrulewidth{0.5pt}

\begin{abstract}
The recent direct detection of gravitational waves (GWs) from binary black hole
mergers~\cite{Abbott:2016blz,Abbott:2016nmj} opens up an entirely new non-electromagnetic
window into the Universe making it possible to probe physics that has been hidden or dark to electromagnetic
observations. In addition to cataclysmic events involving black holes, GWs can be
triggered by physical processes and systems involving neutron stars. Properties of neutron stars
are largely determined by the equation of state (EOS)
of neutron-rich matter, which is the major ingredient in calculating the stellar
structure and properties of related phenomena, such as gravitational wave emission
from elliptically deformed pulsars and neutron star binaries. Although the EOS of neutron-rich
matter is still rather uncertain mainly due to the poorly known density dependence of
nuclear symmetry energy at high densities, significant progress has been made
recently in constraining the symmetry energy using data from terrestrial nuclear
laboratories. These constraints could provide useful information on the limits of GWs expected from neutron stars.
Here after briefly reviewing our previous work on constraining gravitational radiation from elliptically deformed pulsars with terrestrial nuclear laboratory data in light of the recent gravitational
wave detection, we estimate the maximum gravitational wave strain amplitude, using an optimistic value for the
breaking strain of the neutron star crust, for 15 pulsars at distances 0.16 kpc to 0.91 kpc
from Earth, and find it to be in the range of $\sim[0.2-31.1]\times 10^{-24}$, depending on the details of the EOS
used to compute the neutron star properties. Implications are discussed.

\end{abstract}

\section{Introduction}

Gravitational waves are tiny disturbances in space-time predicted by General Relativity
in 1916. A century after the fundamental predictions of Albert Einstein, the Laser Interferometer
Gravitational-Wave Observatory (LIGO) has detected directly gravitational waves from black
hole mergers~\cite{Abbott:2016blz,Abbott:2016nmj}. This detection has the potential to transform profoundly
our understanding of the Universe, as gravitational wave astronomy opens up the possibility to
probe physics that has been hidden to current electromagnetic observations~\cite{Maggiore:2007}.
Because gravity interacts extremely weakly with matter, gravitational waves carry much cleaner
and detailed picture of their sources as opposed to their electromagnetic
counterparts~\cite{Flanagan:2005yc}.

Several types of gravitational waves and their corresponding sources have been discussed in the literature:
(i) {\it Inspiral gravitational waves} are triggered during the latest stages of compact binary systems
where the two objects finally collide and merge. These systems typically consist of two neutron stars, two
black holes, as demonstrated by LIGO~\cite{Abbott:2016blz,Abbott:2016nmj}, or a neutron star and a black hole whose orbit decayed
to the point that the two masses are about to coalesce. This requires one of the original stars to be massive enough to
undergo collapse to a compact object without destroying its companion, and without disrupting the bound orbit~\cite{Riles:2012yw}.
(ii) {\it Burst gravitational waves} are generated in sudden cataclysmic events, such as supernova explosions
or gamma-ray bursts~\cite{Riles:2012yw}. A spherically symmetric explosion cannot lead to emission of GWs in General Relativity.
In order to produce GWs a supernova must exhibit some asymmetry. It is known that many pulsars formed in
supernovae have large speeds relative to neighboring stars (high "birth kicks"), and this suggests strongly that
certain spernovae do exhibit considerable non-spherical motion~\cite{Ott:2006qp}. (iii) {\it Stochastic gravitational waves} could
even have been produced during the very early Universe, well before any stars had been formed, merely as a consequence of
the dynamics and expansion of the Cosmos~\cite{Riles:2012yw}. Very similar to the Cosmic Micro-wave Background (CMB), which is an
electromagnetic leftover from the Big Bang, these GWs originate from many independent events amounting to a Cosmic Gravitational-wave
Background (CGB). Such gravitational waves may carry critical information about the very early Universe, as they would have been
stretched out as the Cosmos expanded. (iv) {\it Continuous gravitational waves} are generated by sources with steady and well-defined
frequency. Examples of such sources are binary neutron star or black hole systems orbiting the common center of
mass, or a (rapidly) spinning neutron star with some long-living axial asymmetry~\cite{Jaranowski:1998qm}.

Ground-based gravitational wave observatories, such as LIGO and VIRGO, will also allow studies of a large population of neutron stars. Additionally, in space eLISA(Evolved Laser Interferometric Space Antenna), a planned space mission by the European Space Agency expected to launch in the early 2020s, will provide an unprecedented instrument for GWs search and detection~\cite{AmaroSeoane:2012je}. The eLISA will survey the low-frequency gravitational-wave bandwidth and is expected to detect a wide variety of events and systems in Space.

Rotating neutron stars are considered to be one of the main prospects for sources of continuous GWs potentially
detectable by the LIGO~\cite{Abbott:2004ig} and VIRGO (e.g. Ref.~\cite{Acernese:2007zzb}) laser interferometric
observatories. According to General Relativity, a perfectly symmetric rotating object self-bound by gravity does not
generate GWs. To emit GWs over an extended time period, a pulsar needs to exhibit a long-living axial asymmetry,
e.g., a "mountain" on its surface~\cite{Jaranowski:1998qm}. In the literature, several different mechanisms causing such
asymmetries have been discussed:
\begin{enumerate}[label=(\roman*)]
  \item{Anisotropic stress built up during the crystallization period of the neutron star crust may be able
  to support long term asymmetries, such as static "mountains" on the neutron star surface~\cite{PPS:1976ApJ}.}
  \item{In addition, because of its violent formation in supernova the rotational axis of a neutron star may not
  necessarily be aligned with its principal moment of inertia axis, which results in a neutron star
  precession~\cite{ZS:1979PRD}. Because of this, even if the pulsar remains symmetric with respect to its rotational
  axis, it generates GWs~\cite{ZS:1979PRD,Z:1980PRD}.}
  \item{Also, neutron stars have extremely strong magnetic fields, which could create magnetic pressure and,
  in turn, deform the pulsar, if the magnetic and rotational axes do not coincide~\cite{BG:1996AA}.}
\end{enumerate}
These mechanisms generally cause a triaxial pulsar configuration. GWs are characterized by a very small dimensionless strain
amplitude, $h_0$. The magnitude of the strain amplitude depends on how much the pulsar is distorted from axial symmetry which depends on details of the EOS of neutron-rich matter. The EOS of nuclear matter under extreme pressure, density and/or isospin asymmetry is still largely uncertain and rather theoretically controversial. A major source of these uncertainties is the rather poorly known density dependence of the nuclear symmetry energy, $E_{sym}(\rho)$ which encodes the information about the energy associated with the neutron-proton asymmetry in neutron-rich matter, see, e.g.~\cite{Li:1997px,ibook01,Lattimer:2004pg,Bar05,Ste05,EPJA}. Fortunately, both nuclear structures and reactions provide useful means to probe the $E_{sym}(\rho)$ in terrestrial nuclear laboratories \cite{Bah14}. For example, several observables in heavy-ion reactions are known to be sensitive to the $E_{sym}(\rho)$ from sub-saturation to supra-saturation densities, see, e.g., Refs.~\cite{Li97,Li00,Li02,Rei16}. For a comprehensive recent review, see articles in Ref.~\cite{EPJA}. Thanks to the hard work of many people in both nuclear physics and astrophysical communities, considerable progress has been achieved in recent years in constraining the symmetry energy especially around and below the nuclear matter saturation density using data from both terrestrial experiments and astrophysical observations, see, e.g., Refs.~\cite{LCK08,Tsa12,LiHan13,Hor14,Bal16}. In this chapter, we first briefly review our earlier work on constraining the GWs expected from elliptically deformed pulsars~\cite{Krastev:2008PLB} in light of the recent direct gravitational wave detections~\cite{Abbott:2016blz,Abbott:2016nmj}. Applying several nucleonic EOSs constrained by laboratory nuclear experiments, we then estimate the GW strain amplitude for fifteen pulsars for selected neutron star configurations. Our focus is on illustrating effects of nuclear symmetry energy on the GW strain amplitude from deformed pulsars. Recent reviews on effects of the density dependence of nuclear symmetry energy on GWs from various neutron star oscillations~\cite{New14}, binary neutron star mergers~\cite{Fattoyev:2013rga}, the pure general relativistic w-mode~\cite{Wen09} as well as the EOS-gravity degeneracy~\cite{He15} can be found in the referred references.

\section{Strain amplitude of gravitational waves from deformed pulsars}

For consistency, we first recall the formalism applied to compute the gravitational wave strain amplitude following closely the discussion of Ref.~\cite{Krastev:2008PLB}. A rotating neutron star generates GWs if it has
some long-living axial asymmetry. As already discussed in the introduction, there are several mechanisms that could
lead to stellar deformations, and in turn gravitational wave emission. Generally, such processes result in triaxial
neutron star configuration, which in the quadrupole approximation, would generate GWs at {\it twice} the rotational
frequency of the star~\cite{Abbott:2004ig}. These waves are characterized by a strain amplitude at the Earth's
vicinity given by~\cite{HAJS:2007PRL}
\begin{equation}\label{Eq.1}
h_0=\frac{16\pi^2G}{c^4}\frac{\epsilon I_{zz}\nu^2}{r}.
\end{equation}
In the above equation $\nu$ is the neutron star rotational frequency, $I_{zz}$ is the stellar principal moment of inertia,
$\epsilon=(I_{xx}-I_{yy})/I_{zz} $ is its equatorial ellipticity, and $r$ is the distance to Earth. The
ellipticity is related to the maximum quadrupole moment of the star via~\cite{Owen:2005PRL}
\begin{equation}\label{Eq.2}
\epsilon = \sqrt{\frac{8\pi}{15}}\frac{\Phi_{22}}{I_{zz}},
\end{equation}
where for {\it slowly} rotating (and static) neutron stars
$\Phi_{22}$ can be written as~\cite{Owen:2005PRL}
\begin{equation}\label{Eq.3}
\Phi_{22,max}=2.4\times
10^{38}g\hspace{1mm}cm^2\left(\frac{{\sigma}_{max}}{10^{-2}}\right)\left(\frac{R}{10km}\right)^{6.26}
\left(\frac{1.4M_{\odot}}{M}\right)^{1.2}.
\end{equation}
In this expression ${\sigma}_{max}$ is the breaking strain of the neutron star crust which is rather uncertain at present time.
Although earlier studies estimated the value of the breaking strain to be in the range ${\sigma}_{max}=[10^{-5}-10^{-2}]$~\cite{HAJS:2007PRL},
more recent investigations using molecular dynamics suggested that the breaking strain could be as large as
${\sigma}_{max}=0.1$~\cite{Horowitz:2009ya}. Using a rather conservative value of
${\sigma}_{max}=0.01$ in a previous work~\cite{Krastev:2008PLB} we reported the nuclear constraints on the gravitational wave signals to be
expected from several pulsars close to Earth. Here, to estimate the maximum value of $h_0$, we take ${\sigma}_{max}=0.1$ and revisit some key issues. From Eqs.~(\ref{Eq.1}) and (\ref{Eq.2}) it is clear that $h_0$ does not depend on the moment of inertia
$I_{zz}$, and the total dependence upon the underlying EOS is carried by the quadrupole moment $\Phi_{22}$. Thus Eq.~(\ref{Eq.1}) can be
rewritten as
\begin{equation}\label{Eq.4}
h_0=\chi\frac{\Phi_{22,max}\nu^2}{r},
\end{equation}
with $\chi=\sqrt{2048\pi^5/15}G/c^4$.

In a previous work~\cite{WKL:2008ApJ} we have computed the moment of inertia of both static and rotating neutron stars.
In the case of slowly rotating neutron stars, in Ref.~\cite{Lattimer:2005}, it has been shown that the following empirical relation
\begin{equation}\label{Eq.5}
I\approx 0.237MR^2\left[1+4.2\left(\frac{Mkm}{M_{\odot}R}\right)+90\left(\frac{Mkm}{M_{\odot}R}\right)^4\right]
\end{equation}
holds for neutron star configurations with masses greater than $1M_{\odot}$ and a wide range of EOSs.
We first solve the Tolman-Oppenheimer-Volkoff (TOV) equations
(TOV)~\cite{Tolman:1939jz,PhysRev.55.374}
\begin{equation}\label{Eq.6}
\frac{dP(r)}{dr}=-\frac{\epsilon(r)m(r)}{r^2}
\left[1+\frac{P(r)}{\epsilon(r)}\right]
\left[1+\frac{4\pi{r^3}p(r)}{m(r)}\right]
\left[1-\frac{2m(r)}{r}\right]^{-1},
\end{equation}
\begin{equation}\label{Eq.7}
\frac{dm(r)}{dr}=4\pi\epsilon(r)r^{2},
\end{equation}
to compute the neutron star mass, $M$, and radius, $R$. We then proceed to calculate the moment of inertia and the quadrupole moment
using Eq.~(\ref{Eq.5}) and Eq.~(\ref{Eq.3}) respectively, and the neutron star ellipticity, $\epsilon$, via Eq.~(\ref{Eq.2}). For
rotational frequencies up to $\sim 300Hz$, global properties of rotating neutron stars stay approximately constant~\cite{WKL:2008ApJ}.
Therefore for slowly rotating stars, if one knows the pulsar's rotational frequency and its distance from the Earth, the above formalism
can be applied to estimate the GW strain amplitude. Then these estimates can be compared with the current upper limits for the sensitivity
of the laser interferometric observatories, such as LIGO and VIRGO~\cite{Krastev:2008PLB}.
\begin{figure}[!t]
\centering
\includegraphics[totalheight=4.0in]{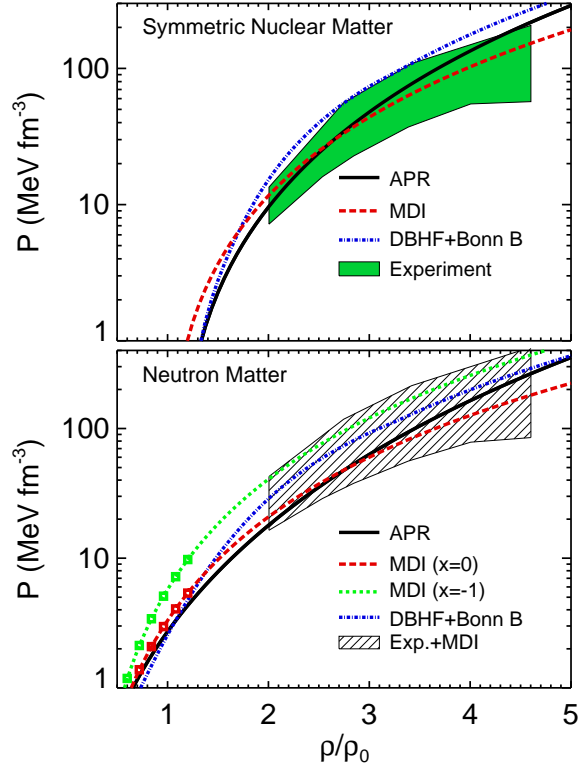}
\vspace{5mm} \caption{(Color online) Pressure as a function of
density for symmetric (upper panel) and pure neutron (lower panel)
matter. The green area in the upper panel is the experimental
constraint on symmetric matter extracted by Danielewicz, Lacey and
Lynch~\cite{Danielewicz:2002pu} from analyzing the collective flow in
relativistic heavy-ion collisions. The corresponding constraint on
the pressure of pure neutron matter, obtained by combining the
flow data and an extrapolation of the symmetry energy functionals
constrained below $1.2\rho_0$ ($\rho_0=0.16 fm^{-3}$) by the
isospin diffusion data, is the shaded black area in the lower
panel. Taken from Ref.~\cite{Krastev:2008PLB}.}\label{f1}
\end{figure}

\section{Model EOSs constrained by terrestrial nuclear experiments}

\begin{figure}[t!]
\centering
\includegraphics[scale=0.9]{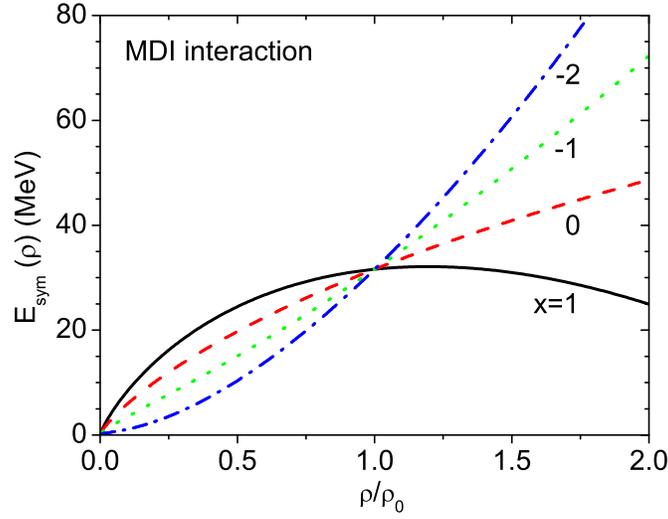}
\caption{(Color online) The density dependence of the nuclear symmetry energy for
different values of the parameter $x$ in the MDI interaction.
Taken from Ref.~\cite{Li:2004cq}.} \label{f1b}
\end{figure}

For completeness, below we summarize relevant EOS details following closely Ref.\cite{Krastev:2008PLB}. We assume a simple model of stellar matter of nucleons and light
leptons (electrons and muons) in beta-equilibrium. For many astrophysical studies, it is more convenient to express the EOS in terms of the pressure as a function of density $\rho$ and isospin asymmetry $\delta=(\rho_n-\rho_p)/\rho$ where $\rho_n$ and $\rho_p$ are the densities of neutrons and protons.
In Fig.~\ref{f1} we show the pressure as a function of density for two
extreme cases: symmetric (upper panel) and pure neutron matter (lower panel). We pay particular attention to the EOS
calculated with the MDI~\cite{Das:2002fr,Li04} (momentum-dependent) interaction because its symmetry energy has been constrained
in the subsaturation density region by the available nuclear laboratory data~\cite{Tsang04}. Here we emphasize that the EOS of symmetric nuclear matter with
the MDI interaction is constrained by the available data on collective flow in relativistic heavy-ion collisions~\cite{Danielewicz:2002pu}. The single-particle
potential corresponding to the MDI EOS is given by
\begin{eqnarray}
&&U_{\tau }(\rho ,T,\delta ,\vec{p},x)=A_{u}(x)\frac{\rho _{-\tau
}}{\rho
_{0}}+A_{l}(x)\frac{\rho _{\tau }}{\rho _{0}}  \nonumber  \label{mdi} \\
&&+B\left( \frac{\rho }{\rho _{0}}\right) ^{\sigma }(1-x\delta ^{2})-8\tau x%
\frac{B}{\sigma +1}\frac{\rho ^{\sigma -1}}{\rho _{0}^{\sigma
}}\delta \rho
_{-\tau }  \nonumber \\
&&+\sum_{t=\tau ,-\tau }\frac{2C_{\tau ,t}}{\rho _{0}}\int d^{3}\vec{p}%
^{\prime }\frac{f_{t}(\vec{r},\vec{p}^{\prime
})}{1+(\vec{p}-\vec{p}^{\prime })^{2}/\Lambda ^{2}},  \label{MDIU}
\end{eqnarray}
where $\tau =1/2$ ($-1/2$) for neutrons (protons), $x$, $A_{u}(x)$, $A_{\ell }(x)$, $B$, $C_{\tau ,\tau }$,$C_{\tau ,-\tau }$, $\sigma
$, and $\Lambda $ are parameters discussed in Ref.~\cite{Das:2002fr}. The parameter $x$ is introduced in Eq. (\ref{MDIU}) to account
for the largely uncertain density behavior of the nuclear symmetry energy $E_{sym}(\rho)$ as predicted by various models of the
nuclear interaction and/or many-body approaches. As illustrated in Fig.~\ref{f1b}, the different values of $x$ result in a wide range
of possible behaviors for the density dependence of the nuclear symmetry energy. It was demonstrated that only EOSs with $x$ in the range between -1 and 0 have symmetry energies consistent with the
isospin-diffusion laboratory data~\cite{Li:2005jy} and the fiducial value for the neutron-skin of $^{208}Pb$~\cite{Li:2005sr}. We therefore consider only these two limiting cases in
calculating the boundaries of the possible neutron star configurations.
The green area in Fig.~\ref{f1}, in the density range of $\rho_0=[2.0-4.6]$, represents the experimental constraint on the pressure of symmetric nuclear matter, $P_0$, extracted from analysis of collective flow data from relativistic heavy-ion collisions~\cite{Danielewicz:2002pu}.
The pressure of pure neutron matter, $P_{PNM}=P_0+\rho^2dE_{sym}(\rho)/d\rho$, is determined by the density dependence of the nuclear symmetry energy $E_{sym}(\rho)$. Because the constraints on the nuclear symmetry energy are available only for densities lower than approximately $1.2\rho_0$,
as shown by the red and green square symbols in the lower panel of Fig.~\ref{f1}, the most reliable estimate of the EOS of neutron-rich
matter can be therefore computed by extrapolating the model of the EOS for symmetric nuclear matter and the symmetry energy to higher densities.
In the lower panel, the black shaded region represents the best estimate on pressure of neutron matter at high densities predicted by
the MDI interaction with $x=0$ and $x=-1$ as the lower and upper bounds on the symmetry energy and the constrained EOS of symmetric matter
with flow data. The estimated error bars of the neutron matter EOS at higher densities are much wider compared to the uncertainty range
of the EOS of symmetric matter. This is consistent with the estimate in Ref.~\cite{Danielewicz:2002pu}. In Fig.~\ref{f1}, in addition to the
MDI EOS, we also display results by Akmal et al.~\cite{Akmal:1998cf} with the $A18+\delta\upsilon+UIX*$ interaction (APR) and Dirac-Brueckner-Hartree-Fock (DBHF) calculations~\cite{Alonso:2003aq,Krastev:2006ii} with Bonn B One-Boson-Exchange (OBE) potential (DBHF+Bonn B)~\cite{Machleidt:1989}.
Table~\ref{tab.1} summarizes the saturation properties of the EOSs used in this work.
\begin{table}[!t]
\caption{Saturation properties of the nuclear EOSs (for symmetric
nuclear matter) shown in Fig.~1.}
\begin{center}
\begin{tabular}{lccccc}\label{tab.1}
EOS &  $\rho_0$ & $E_s$ & $\kappa$ & $m^*(\rho_0)$ & $E_{sym}(\rho_0)$ \\
    &    $(fm^{-3})$ &  $(MeV)$  &  $(MeV)$   & $(MeV/c^2)$ &
    $(MeV)$ \\
\hline\hline
MDI         & 0.160 & -16.08 & 211.00 & 629.08 &  31.62 \\
APR         & 0.160 & -16.00 & 266.00 & 657.25 &  32.60 \\
DBHF+Bonn B & 0.185 & -16.14 & 259.04 & 610.30 &  33.71 \\
\hline
\end{tabular}
\end{center}
\vspace{3mm}{\small The first column identifies the equation of
state. The remaining columns exhibit the following quantities at
the nuclear saturation density: saturation (baryon) density;
energy-per-particle; compression modulus; nucleon effective mass;
symmetry energy. Taken from Ref.~\cite{Krastev:2008PLB}.}
\end{table}

For modeling neutron stars, at baryon densities below approximately $0.07fm^{-3}$ the EOSs used in this work are supplemented by a crustal EOS, which is more applicable at lower densities. For the inner crust we apply the EOS by Pethick et al.~\cite{PRL1995} and
for the outer crust the one by  Haensel and Pichon~\cite{HP1994}. At higher densities we assume a continuous functional for the equations of
state applied here. (For details of the extrapolation procedure of the DBHF+Bonn B EOS see Ref.~\cite{Krastev:2006ii}.)
The MDI EOS has been used in various calculations of neutron star properties
and related phenomena. For example, it has been used to constrain the neutron star radius~\cite{Li:2005sr} with an estimated range consistent with the observational data~\cite{Hes06}. It has been also applied to constrain possible time variations of the gravitational constant $G$~\cite{Krastev:2007en} with the help of
the {\it gravitochemical heating} approach developed by Jofre et al.~\cite{Jofre:2006ug}. In addition, we applied the MDI EOS to limit a number of other global, transport and thermal properties of both static and rapidly rotating neutron stars~\cite{WKL:2008ApJ,KLW2,Newton,Junxu}.
\begin{figure}[!b]
\vspace{5mm}
\centering
\includegraphics[totalheight=5.0cm]{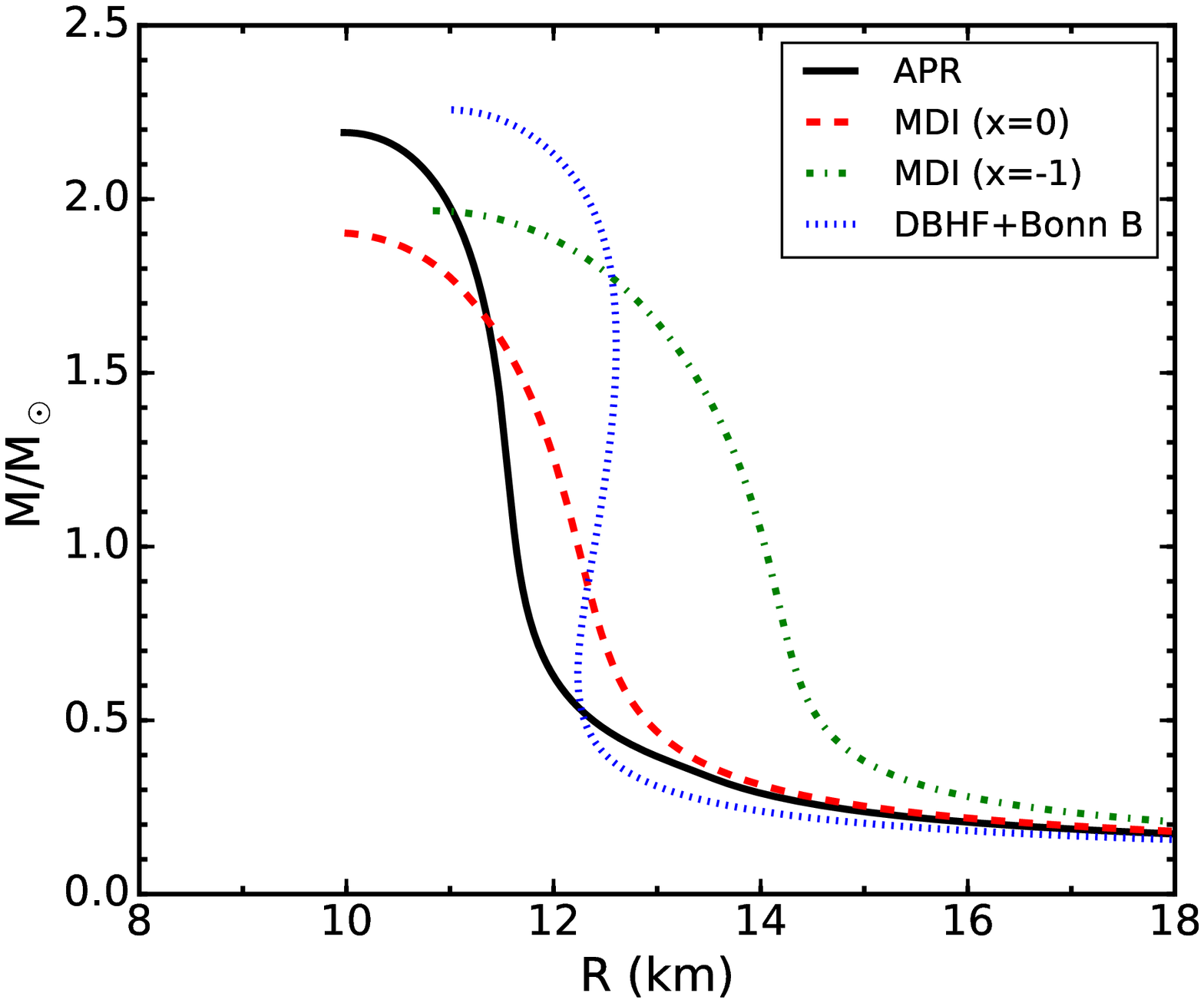}
\hspace{5mm}
\includegraphics[totalheight=5.0cm]{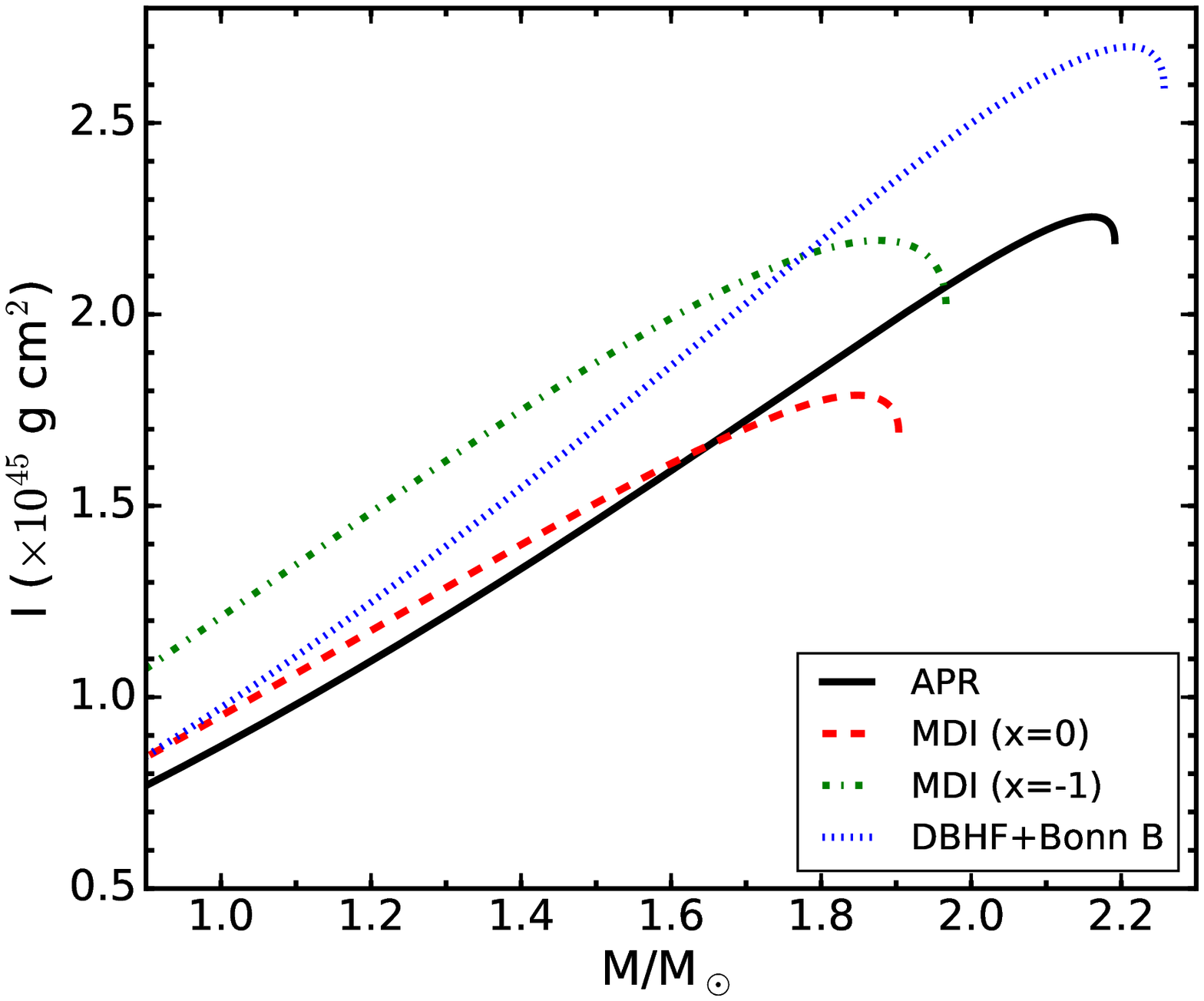}
\vspace{5mm} \caption{(Color online) Neutron star mass-radius relation (left panel) and
moment of inertia (right panel). Right panel adapted from Ref.~\cite{WKL:2008ApJ}.}\label{f2}
\end{figure}

\section{Constraining properties of gravitational waves from deformed pulsars}

Fig.~\ref{f2} displays the neutron star mass-radius relation (left frame), where $M$ and $R$ are calculated
by integrating Eqs. (\ref{Eq.6}) and (\ref{Eq.7}), and moment of inertia (right frame). The moment of inertia is calculated
through Eq.~(\ref{Eq.5}), which as already mentioned holds for slowly rotating neutron stars.
\begin{figure}[!t]
\vspace{5mm}
\centering
\includegraphics[totalheight=5.0cm]{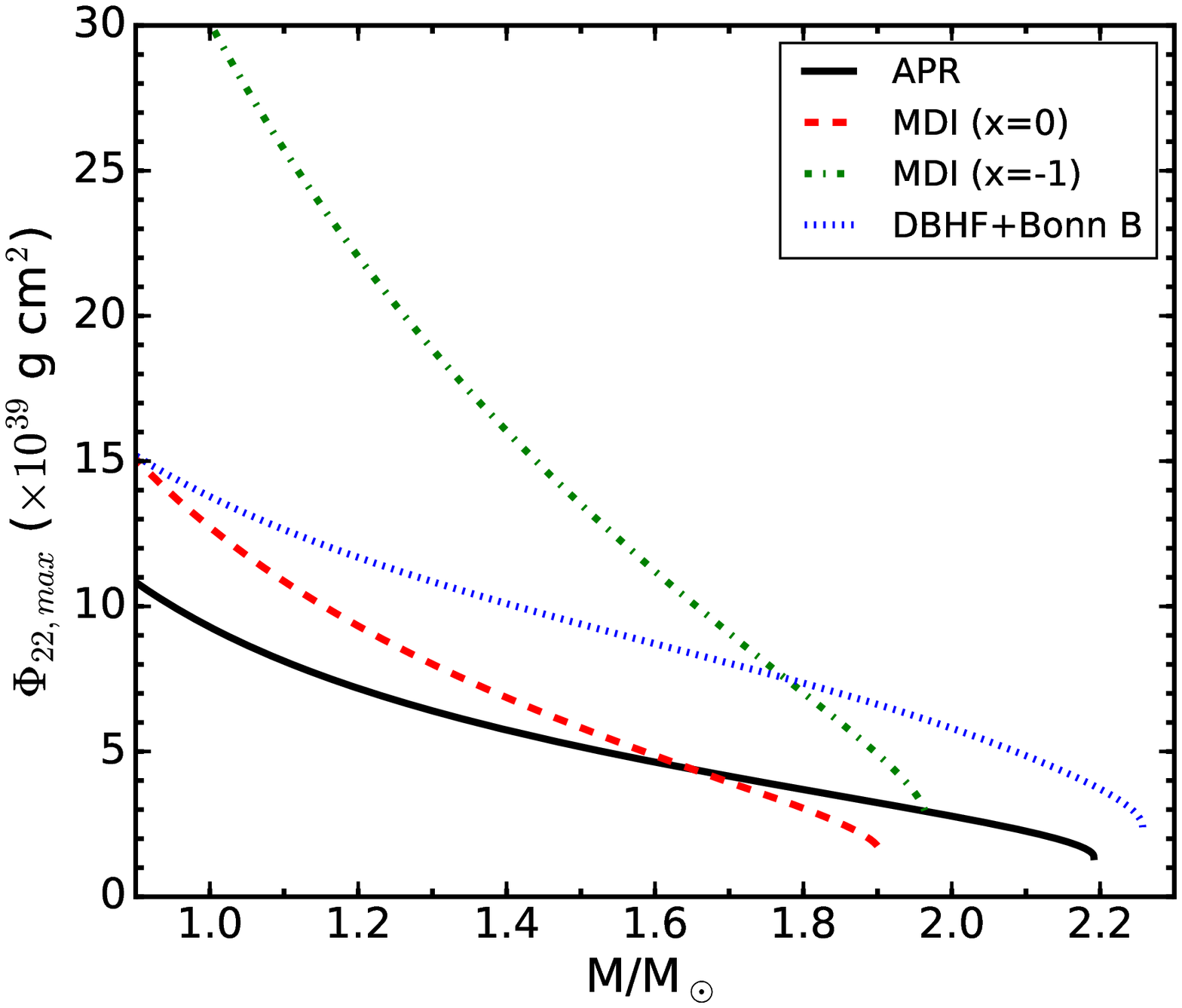}
\hspace{5mm}
\includegraphics[totalheight=5.0cm]{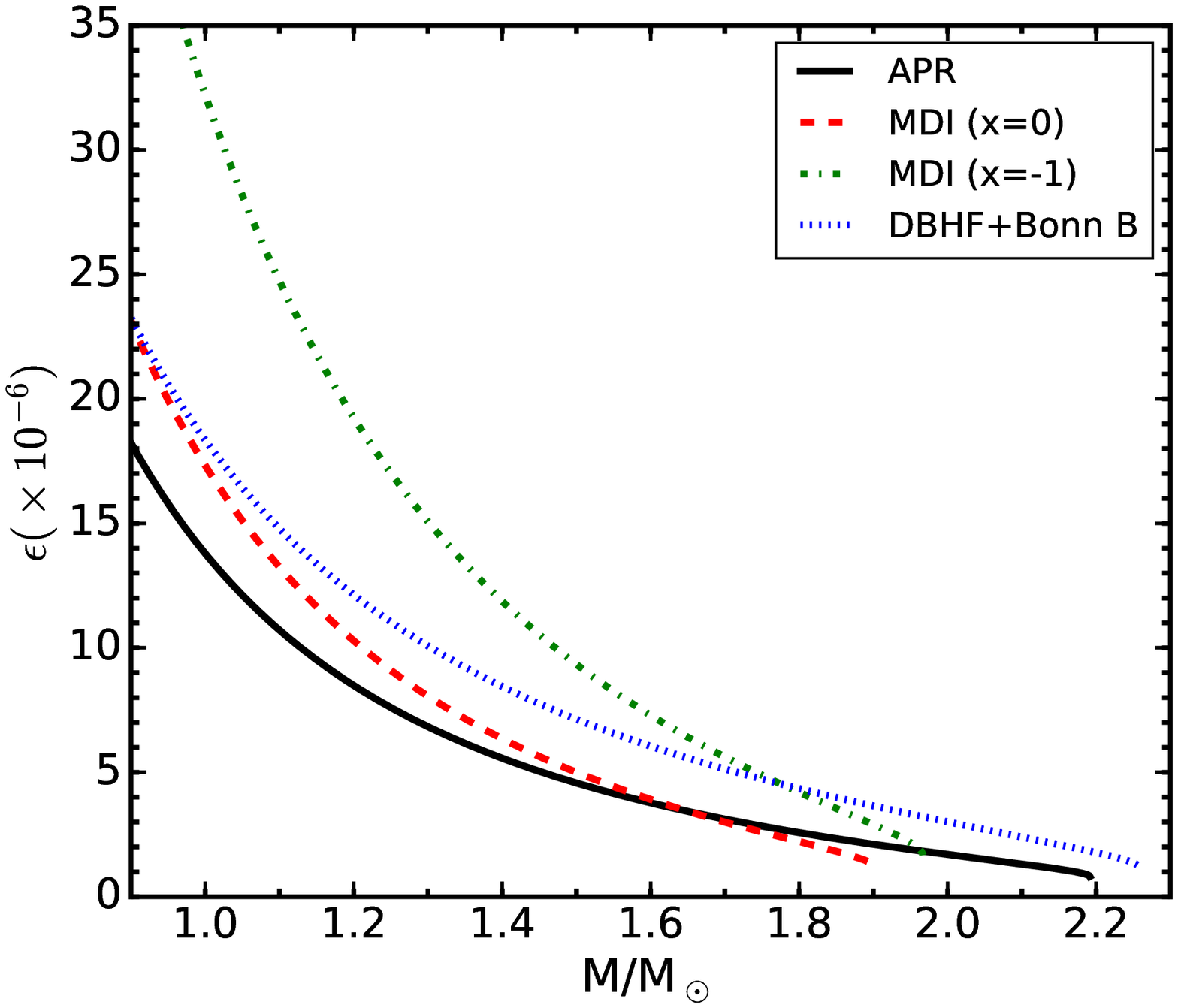}
\vspace{5mm} \caption{(Color online) Neutron star quadrupole moment (left panel)
and ellipticity (right panel). Adapted from Ref.~\cite{Krastev:2008PLB}.}\label{f3}
\end{figure}
In Fig.~\ref{f3} we display the neutron star quadrupole moment (left panel) and ellipticity (right panel)
as computed through Eqs.~(\ref{Eq.3}) and (\ref{Eq.2}) respectively. The quadrupole moment decreases with increasing
neutron star mass for the four equations of state applied in this work. As noted previously in Ref.~\cite{Krastev:2008PLB},
this decrease is dependent upon the EOS and is most pronounced for the MDI EOS with $x=-1$. This trend is explained in terms
of increasing the central density with the neutron star mass. Heavier pulsars have higher central density and because $\Phi_{22}$
measures the pulsar's degree of distortion (Fig.~\ref{f2}, left panel), they also exhibit smaller deformations compared to
stars with lower central densities. In addition, the neutron star mass is well known to be mainly determined by the symmetric
component of the nuclear equation of state. On the other hand, the neutron star radius depends strongly on the slope of the
nuclear symmetry energy. In other words, an EOS with a stiff $E_{sym}$, such as the MDI EOS with $x=-1$, produces less centrally
condensed neutron star models, and in turn pulsars with a greater degree of deformation. In this respect, we recall that the MDI
equation of state with $x=-1$ results in stellar models with greater radii compared to those of configurations calculated
with the rest of the EOS applied in this work (e.g., Eq.~(\ref{Eq.2})). These findings are consistent with previous investigations
suggesting that more centrally condensed stellar configurations are less deformed by rapid rotation~\cite{FPI:1984Nature}.
As concluded in Ref.~\cite{Krastev:2008PLB}, such neutron star models are also to be expected to be  more ``resistant'' to
any kind of distortion.

\begin{figure}[!t]
\vspace{12mm}
\centering
\includegraphics[totalheight=4.5in]{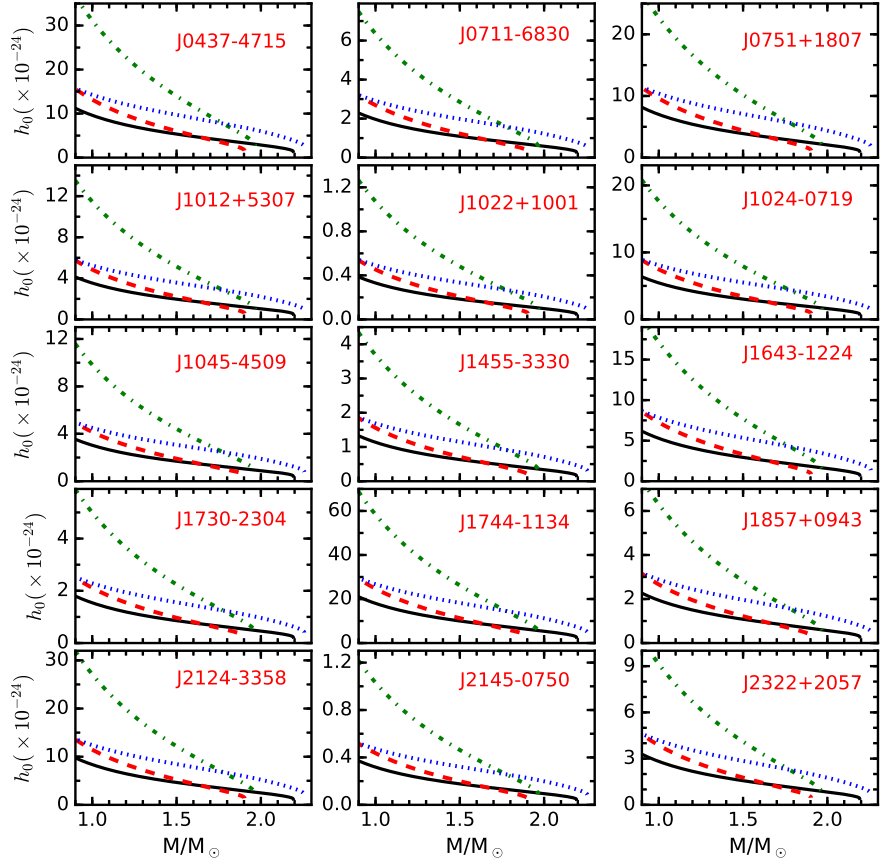}
\vspace{5mm} \caption{(Color online) Gravitational wave strain amplitude as a function of the neutron star mass.
The figure legend is the same as in Figs.~\ref{f2} and \ref{f3}.}\label{f4}
\end{figure}

Because the ellipticity $\epsilon$ is proportional to $\Phi_{22}$, it also decreases with the increase of the neutron star mass (see
the right panel of Fig~\ref{f3}). It is seen that models with stiff
symmetry energy, such as the MDI EOS with $x=-1$, favor larger crust "mountains". These results are consistent with recent investigations~\cite{Fattoyev:2013rga}, involving one of us, where it was demonstrated that gravitational wave signals could provide critical information about the high-density behavior of nuclear symmetry energy.

In Fig.~\ref{f4} we show the gravitational wave strain amplitude, $h_0$, versus stellar mass for selected pulsars. We chose
fifteen neutron stars with rotational frequencies less than 300 Hz so that $I_{zz}$ and $\Phi_{22}$ can be approximately estimated via
Eqs. (\ref{Eq.5}) and (\ref{Eq.3}). These pulsars are also relatively close to Earth ($r < 1$ kpc) to maximize $h_0$ of the expected GWs.
The results shown in Fig.~\ref{f4} demonstrate that the strain amplitude depends on the details of the EOS of neutron-rich matter and
the dependence is more pronounced for lighter stellar models. The dependence upon the EOS is also stronger for configurations
calculated with stiffer equations of state. As already discussed, such models have lower central densities and are therefore less bound
by gravity. This makes them more easily distorted by various mechanisms (see introduction section) and as a result greater prospects
for sources of stronger GWs (see Eq.~(\ref{Eq.1})). The gravitational wave strain amplitude calculated with the MDI EOS with
$x=0$ and $x=-1$ serve as a limit on the maximum $h_0$ of the gravitational waves generated by these neutron stars. Here we recall that
these estimates ignore the distance measurement uncertainties, and they also should be considered as upper limits of $h_0$ because $\Phi_{22}$
has been calculated with $\sigma=0.1$, and it may go as low as $10^{-5}$.

\begin{table}[!t]
\begin{center}
\caption{Properties of nearby pulsars considered in this study.}\vspace{3mm}
\begin{tabular}{ccccccc}\label{tab.2}
PSR        & $\nu$  & $r$      & $h_0^{(APR)}$ & $h_0^{(x=0)}$ & $h_0^{(x=-1)}$ &  $h_0^{(DBHF+Bonn B)}$ \\
           & $(Hz)$ & $(kpc) $ &               &                &               &                        \\
\hline\hline
J0437-4715  & 173.91 & 0.16 & 5.937   &  7.092  & 16.582  &  10.439 \\
J0711-6830  & 182.12 & 0.86 & 1.211   &  1.447  & 3.383   &  2.130  \\
J0751+1807  & 287.46 & 0.60 & 4.325   &  5.167  & 12.081  &  7.606  \\
J1012+5307  & 190.27 & 0.52 & 2.186   &  2.612  & 6.107   &  3.845  \\
J1022+1001  &  62.50 & 0.60 & 0.204   &  0.244  & 0.571   &  0.360  \\
J1024-0719  & 193.72 & 0.35 & 3.367   &  4.023  & 9.406   &  5.921  \\
J1045-4509  & 133.80 & 0.30 & 1.874   &  2.239  & 5.235   &  3.296  \\
J1455-3330  & 125.16 & 0.70 & 0.703   &  0.840  & 1.963   &  1.236  \\
J1643-1224  & 216.36 & 0.45 & 3.267   &  3.903  & 9.125   &  5.745  \\
J1730-2304  & 123.11 & 0.50 & 0.952   &  1.137  & 2.659   &  1.674  \\
J1744-1134  & 245.43 & 0.17 & 11.128  &  13.294 & 31.082  &  19.568 \\
J1857+0943  & 186.49 & 0.91 & 1.200   &  1.434  & 3.353   &  2.111  \\
J2124-3358  & 202.79 & 0.25 & 5.166   &  6.172  & 14.430  &  9.084  \\
J2145-0750  &  62.30 & 0.62 & 0.197   &  0.235  & 0.549   &  0.346  \\
J2322+2057  & 207.97 & 0.78 & 1.741   &  2.080  & 4.864   &  3.062  \\
\hline
\end{tabular}
\end{center}
\vspace{3mm} {\small The first column identifies the pulsar. The remaining columns are the rotational frequency, distance to Earth, and the calculated upper limits on the gravitational wave strain amplitude, for the equations of state applied in this study. Since the masses of most of these pulsars are
presently unknown we have used a canonical value of $1.4M_{\odot}$ to compute the gravitational-wave strain amplitude, $h_0$, which is given in units of
$1.0\times 10^{-24}$.}
\end{table}

The results shown in Fig.~\ref{f4} are summarized in Table~\ref{tab.2} where we include the upper limits on $h_0$ for
the pulsars considered here. The specific results listed in the table are for neutron star configurations of $1.4M_{\odot}$.
Depending on the distance from Earth, rotational frequency, and details of the EOS, the {\it maximal} $h_0$ is estimated to be
in the range $\sim[0.2-31.1]\times 10^{-24}$. The exact values of $h_0$ would depend on the neutron star mass.

\begin{figure}[!t]
\vspace{6mm}
\centering
\includegraphics[totalheight=8.5cm]{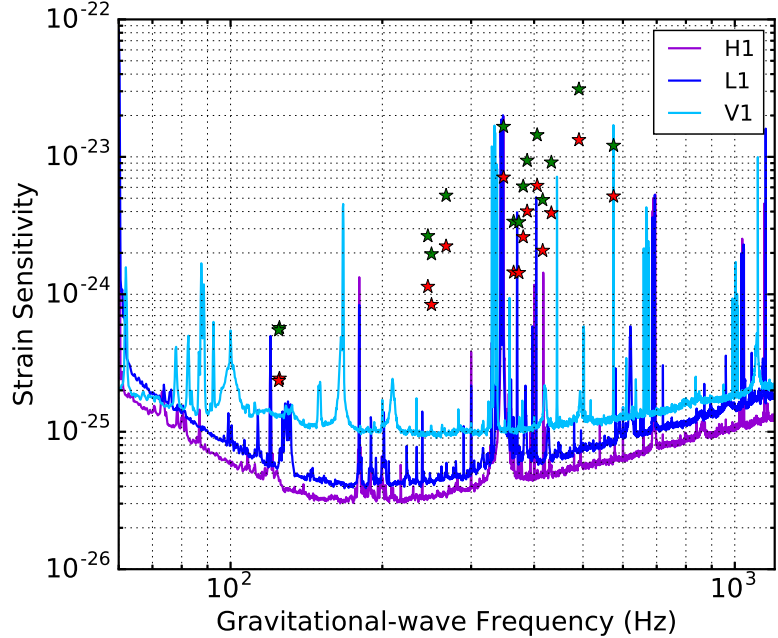}
\vspace{5mm} \caption{(Color online) GW strain amplitude as a function of its frequency from deformed pulsars. The curves give the estimated relative strain sensitivities of the LIGO S6 and VIRGO VSR2/VSR3 runs.The H1, L1, and V1 denote the sensitivities of the LIGO Hanford and Livingston observatories, and VIRGO observatory respectively. The $\star$ symbols show predictions for the GW strain amplitude for models of a $1.4M_{\odot}$ neutron star computed with the $x=0$ (red) and $x=-1$ (green) EOSs. The sensitivity strain data
is taken from the LIGO Open Science Center~\cite{LIGO:Data}.}\label{f5}
\end{figure}

At the end, in Fig.~\ref{f5} we take a different view of the results displayed in Fig.~\ref{f4}. We show the maximal gravitational
wave strain amplitude as a function of the gravitational wave frequency and compare our estimates with the current
strain sensitivity of the LIGO and VIRGO detectors. The specific case shown in the figure is for neutron star configurations with mass $1.4M_{\odot}$ computed with the MDI EOS with $x=0$ and $x=-1$. Because these equations of state are constrained with data from terrestrial nuclear laboratories
they limit the possible neutron star models and therefore the GWs expected from them. The results in Fig.~\ref{f5} (and Table~\ref{tab.2}) would suggest that the GWs from the pulsars considered here should be within the current detection range of LIGO and VIRGO. However, no signal detection from any of these targets were reported. The fact that such a detection has not been made yet deserves a few comments. First and most importantly, in calculating the upper limits on $h_0$ we have used the maximum value of the breaking strain of the neutron star crust, $\sigma=0.1$~\cite{Horowitz:2009ya}, which is very
optimistic. Theoretical calculations of $\sigma$ differ by at least four orders of magnitude, where it could take values as low as
$\sigma=10^{-5}$~\cite{HAJS:2007PRL}. In fact, one of the major challenges is to come up with a reasonable mechanism that causes large deformations~\cite{Andersson:2009yt}. In this regard, accretion-powered pulsars could be promising candidates due
to the expected asymmetry of the accretion flow near the stellar surface~\cite{Fattoyev:2013rga}. However, the required computational
modeling of such systems is not trivial because of their complicated dynamics. Second, we have assumed a particular stellar mass of $1.4M_{\odot}$. On
the other hand, as illustrated in Fig.~\ref{f4} $h_0$ decreases with the increase of the neutron star mass which means that heavier pulsars
emit weaker gravitational waves. Although masses of pulsars tend to cluster around the canonical neutron star mass of $1.4M_{\odot}$,
heavier pulsars, with masses $\sim 2M_{\odot}$ have been reported~\cite{Demorest:2010bx,Antoniadis:2013pzd}.
Third, we have applied a very simple approximate relation for the maximum quadrupole moment where we didn't take into account
uncertain effects that may be caused by the properties of the neutron star crust. Lastly, distance estimates based on dispersion
measurements could also be off be a factor of 2-3~\cite{Abadie:2010cf}. We also remind that Eq.~(\ref{Eq.1}) implies that
the best prospects for GWs (from rotating neutron stars) are pulsars relatively close to Earth rotating at high frequencies
($h_0\sim \Phi_{22}\nu^2/r$) -- the results displayed in Figs.~\ref{f4} and \ref{f5} would be modified in favor of GW detection, by increasing
the rotational frequency (and/or decreasing distance to Earth, $r$). At the and, for more realistic calculations,
$\Phi_{22}$ needs to be computed exactly numerically by solving the Einstein field equations for rapidly rotating neutron stars.

\section{Summary and outlook}

In this work we revisited and updated our estimates on the upper limits on the gravitational wave signals expected from
elliptically deformed millisecond pulsars and compared our predictions with the current strain sensitivity of LIGO and VIRGO.
More specifically, we provided estimates on the strain amplitude of gravitational waves to be expected from fifteen
pulsars at distances from Earth 0.16 kpc to 0.91 kpc and rotational frequencies below 300 Hz.
We have applied the MDI EOS constrained by nuclear laboratory data and calculated the upper limit on the
gravitational-wave signal to be expected from these pulsars. Depending on the EOS details the {\it maximum} $h_0$ is found to be in the range of $\sim[0.2-31.1]\times 10^{-24}$. Effects of the density dependence of nuclear symmetry energy are examined.

In the near future, new experiments especially those using high intnsity/energy radioactive beams in terrestrial nuclear laboratories will provide more accurate information about the EOS of dense neutron-rich matter, especially the high density behavior of nuclear symmetry energy. The EOS of neutron-rich matter will be better constrained in a wide density range and most of the studies presented in this work will be refined further. The better constrained EOS will then help predict more accurately properties of pulsars and the gravitational waves emitted by them. Similarly, rapid progress in astrophysical observations of neutron stars and measurements of gravitational wave signals will help us better understand the structure of pulsars and extract more accurately information about the underlying EOS. Ultimately, combining results from terrestrial laboratories and astrophysical observations will help us reach the shared goal of thoroughly understanding the nature of dense neutron-rich nucleonic matter.

\section*{Acknowledgements}
This work is supported in part by the US National Science Foundation under Grant No. PHY-1068022 and
the U.S. Department of Energy Office of Science under Award Number DE-SC0013702.

\label{lastpage-01}


\begin{thebibliography}{}
\bibitem{Abbott:2016blz}
  B.~P.~Abbott {\it et al.} [LIGO Scientific and Virgo Collaborations],
  Phys.\ Rev.\ Lett.\  {\bf 116}, no. 6, 061102 (2016).
\bibitem{Abbott:2016nmj}
  B.~P.~Abbott {\it et al.} [LIGO Scientific and Virgo Collaborations],
  Phys.\ Rev.\ Lett.\  {\bf 116}, no. 24, 241103 (2016).
\bibitem{Maggiore:2007}M.~Maggiore, {\it Nature} {\bf 447}, 651 (2007).
\bibitem{Flanagan:2005yc}
  E.~E.~Flanagan and S.~A.~Hughes,
  {\it New J.\ Phys.\ }  {\bf 7}, 204 (2005).
\bibitem{Riles:2012yw}
  K.~Riles,
  Prog.\ Part.\ Nucl.\ Phys.\  {\bf 68}, 1 (2013).
\bibitem{Ott:2006qp}
  C.~D.~Ott, A.~Burrows, L.~Dessart and E.~Livne,
  Phys.\ Rev.\ Lett.\  {\bf 96}, 201102 (2006).
\bibitem{Jaranowski:1998qm}
  P.~Jaranowski, A.~Krolak and B.~F.~Schutz,
  {\it Phys.\ Rev.\  D} {\bf 58}, 063001 (1998).
\bibitem{AmaroSeoane:2012je}
  P.~Amaro-Seoane {\it et al.},
  Class.\ Quant.\ Grav.\  {\bf 29}, 124016 (2012).
\bibitem{Abbott:2004ig}
  B.~Abbott {\it et al.}  [LIGO Scientific Collaboration],
  {\it Phys.\ Rev.\ Lett.\ }  {\bf 94}, 181103 (2005); {\it Phys.\ Rev.\ D} {\bf 76}, 042001 (2007).
\bibitem{Acernese:2007zzb}
  F.~Acernese {\it et al.},
  {\it Class.\ Quant.\ Grav.\ }  {\bf 24}, S491 (2007).
\bibitem{PPS:1976ApJ}V.~R.~Padharipande, D.~Pines, and R.~A.~Smith, {\it Astrophys. J.} {\bf 208}, 550--566 (1976).
\bibitem{ZS:1979PRD}M.~Zimmermann and E.~Szedenits, {\it Phys. Rev. D} {\bf 20}, 351 (1979).
\bibitem{Z:1980PRD}M.~Zimmermann, {\it Phys. Rev. D} {\bf 21}, 891 (1980).
\bibitem{BG:1996AA}S.~Bonazzola and E.~Gourgoulhon, {\it Astron. Astrophys.} {\bf 312}, 675 (1996).
\bibitem{Li:1997px}B.A. Li, C. M. Ko and W. Bauer, {\it Int. J. Mod. Phys. E } {\bf 7}, 147 (1998).
\bibitem{ibook01} \textit{Isospin Physics in Heavy-Ion Collisions at Intermediate Energies},
Eds. B. A. Li and W. Uuo Schr\"{o}der (Nova Science Publishers, Inc, New York, 2001).
\bibitem{Lattimer:2004pg}J. M. Lattimer and M. Prakash, {\it Science} {\bf 304}, 536 (2004).
\bibitem{Bar05} V. Baran, M. Colonna, V. Greco, and M. Di Toro, Phys. Rep. \textbf{410}, 335 (2005).
\bibitem{Ste05} A.W. Steiner, M. Prakash, J.M. Lattimer, and P.J. Ellis, Phys. Rep. \textbf{411}, 325 (2005).
\bibitem{EPJA}  ``Topical issue on nuclear symmetry energy", Eds., B.A. Li, A. Ramos, G. Verde, and I. Vida\~na, Eur. Phys. J. A {\bf 50}, No. 2, (2014).
\bibitem{Bah14}A.B Balantekin, J. Carlson, D.J. Dean, G.M. Fuller, R.J. Furnstahl, M. Hjorth-Jensen, R.V.F. Janssens, B.A. Li, W. Nazarewicz, F.M. Nunes, W.E. Ormand, S. Reddy, B.M. Sherrill, Modern Physics Letters A{\bf 29} (11), 1430010 (2014).
\bibitem{Li97}B. A. Li, C. M. Ko and Z. Z. Ren, Phys. Rev. Lett. {\bf 78}, 1644 (1997).
\bibitem{Li00}B. A. Li, Phys. Rev. Lett. {\bf 85}, 4221 (2000).
\bibitem{Li02}B. A. Li, Phys. Rev. Lett. {\bf 88},192701 (2002).
\bibitem{Rei16} P.-G. Reinhard, A.S. Umar, P.D. Stevenson, J. Piekarewicz, V.E. Oberacker, J.A. Maruhn,
Phys. Rev. C{\bf 93}, 044618 (2016).
\bibitem{LCK08} B.A. Li, L.W. Chen, and C.M. Ko, Phys. Rep. \textbf{464}, 113 (2008).
\bibitem{Tsa12} B.M. Tsang \textit{et al.}, Phys. Rev. C \textbf{86}, 105803 (2012).
\bibitem{LiHan13} B.A. Li and X. Han, Phys. Lett. \textbf{B727}, 276 (2013).
\bibitem{Hor14} C. J. Horowitz, E. F. Brown, Y. Kim, W. G. Lynch, R. Michaels, A. Ono, J. Piekarewicz,
M. B. Tsang, H. Wolter, J. Phys. G: Nucl. Part. Phys. 41, 093001 (2014).
\bibitem{Bal16} M. Baldo and G. F. Burgio, Progress in Particle and Nuclear Physics (2016) in press, arXiv:1606.08838.
\bibitem{Krastev:2008PLB}Plamen G. Krastev, Bao-An Li, and Aaron Worley, {\it Phys. Lett. B} {\bf 668}, 1 (2008).
\bibitem{New14}William G. Newton, Joshua Hooker, Michael Gearheart, Kyleah Murphy, De-Hua Wen, Farrukh Fattoyev and Bao-An Li, Eur.\ Phys.\ J.\ A {\bf 50}, 41 (2014).
\bibitem{Fattoyev:2013rga}F. J. Fattoyev, J. Carvajal, W. G. Newton and B.A. Li, Phys. Rev. C{\bf 87}, 015806 (2013); F.~J.~Fattoyev, W.~G.~Newton and B.~A.~Li,
  Eur.\ Phys.\ J.\ A {\bf 50}, 45 (2014).
\bibitem{Wen09}D.H Wen, B.A. Li and P.G. Krastev, Physical Review C{\bf 80}, 025801 (2009);
W.K. Lin, B.A. Li, J. Xu and C.M. Ko, Physical Review C{\bf 83}, 045802 (2011).
\bibitem{He15} X.T. He, F. J. Fattoyev, B.A. Li and W. G. Newton, Phys. Rev. C{\bf 91}, 015810 (2015).
\bibitem{HAJS:2007PRL}B. Haskell, N. Andersson, D. I. Jones, and L. Samuelsson, {\it Phys. Rev. Lett.} {\bf 99}, 231101 (2007).
\bibitem{Owen:2005PRL}B. J. Owen, {\it Phys. Rev. Lett.} {\bf 95}, 211101 (2005).
\bibitem{Horowitz:2009ya}
  C.~J.~Horowitz and K.~Kadau,
  Phys.\ Rev.\ Lett.\  {\bf 102}, 191102 (2009).
\bibitem{WKL:2008ApJ}A. Worley, P. G. Krastev, and B.A. Li, {\it Astrophys. J.} {bf 685}, 390 (2008).
\bibitem{Lattimer:2005}J. M. Lattimer and B. F. Schutz, {\it Astrophys. J.} {\bf 629}, 979 (2005).
\bibitem{Tolman:1939jz}R. C. Tolman, {\it Phys. Rev.} {\bf 55}, 364 (1939).
\bibitem{PhysRev.55.374}J. R. Oppenheimer and G. M. Volkoff, {\it Phys. Rev.} {\bf 55}, 374 (1939).
\bibitem{Das:2002fr}C.~B.~Das, S.~D.~Gupta, C.~Gale, and B.A. Li, Phys. Rev. C {\bf 67}, 034611 (2003).
\bibitem{Li04} B.A. Li, C. B. Das, S. Das Gupta, and C. Gale, Phys. Rev. C 69, 011603 (R) (2004); Nucl. Phys. A735, 563 (2004).
\bibitem{Tsang04}M.B. Tsang et al., Phys. Rev. Lett. {\bf 92}, 062701 (2004).
\bibitem{Danielewicz:2002pu}P. Danielewicz, R. Lacey and W. G. Lynch, {\it Science} {\bf 298}, 1592 (2002).
\bibitem{Li:2005jy}L.W Chen, C.M Ko and B.A. Li, Physical Review Letters {\bf 94}, 032701 (2005);
B.A. Li and L.W. Chen, Phys. Rev. C {\bf 72},  064611 (2005).
\bibitem{Li:2005sr}A.W Steiner and B.A. Li, Physical Review C {\bf 72}, 041601 (2005);
B.A. Li and A. W. Steiner, Phys. Lett. B {\bf 642}, 436 (2006).
\bibitem{Akmal:1998cf}A.~Akmal, V.~R.~Pandharipande, D.~G.~Ravenhall, Phys. Rev. C {\bf 58}, 1804 (1998).
\bibitem{Alonso:2003aq}D.~Alonso and F.~Sammarruca, Phys. Rev. C {\bf 67}, 054301 (2003).
\bibitem{Krastev:2006ii}P.~G.~Krastev and F.~Sammarruca, Phys. Rev. C {\bf 74}, 025808 (2006).
\bibitem{Machleidt:1989}R.~Machleidt, Adv. Nucl. Phys. {\bf 19}, 189 (1989).
\bibitem{PRL1995}C.~J.~Pethick, D.~G.~Ravenhall, and C.~P.~Lorenz, Nucl. Phys. A {\bf 584}, 675 (1995).
\bibitem{HP1994}P.~Haensel and B.~Pichon, Astron. Astrophys. {\bf 283}, 313 (1994).
\bibitem{Hes06} J.W.T. Hessels et al., Science {\bf 311}, 1901 (2006).
\bibitem{Krastev:2007en}P.~G.~Krastev, B.A. Li, Phys. Rev. C {\bf 76}, 055804 (2007).
\bibitem{Jofre:2006ug}P.~Jofre, A.~Reisenegger, and R.~Fernandez, Phys. Rev. Lett. {\bf 97}, 131102 (2006).
\bibitem{KLW2}P.~G.~Krastev, B.A. Li, and A.~Worley, Astrophys. J. {\bf 676}, 1170 (2008).
\bibitem{Newton}W.G. Newton and B.A. Li, Phys. Rev. C{\bf 80},065809 (2009); W.G Newton, M. Gearheart, B.A. Li,
The Astrophysical Journal Supplement Series {\bf 204} (1), 9 (2012).
\bibitem{Junxu}Jun Xu, Lie-Wen Chen, Bao-An Li and Hong-Ru Ma, The Astrophys. J {\bf 697}, 1549 (2009);
Jun Xu, Lie-Wen Chen, Che Ming Ko and Bao-An Li, Phys. Rev. C{\bf 81}, 055805 (2010).
\bibitem{FPI:1984Nature}J.~L.~Friedman, L.~Parker, and J.~R~Ipser, Nature {\bf 312}, 25 (1984).
\bibitem{Andersson:2009yt}
  N.~Andersson, V.~Ferrari, D.~I.~Jones, K.~D.~Kokkotas, B.~Krishnan, J.~S.~Read, L.~Rezzolla and B.~Zink,
  Gen.\ Rel.\ Grav.\  {\bf 43}, 409 (2011).
\bibitem{Demorest:2010bx}
  P.~Demorest, T.~Pennucci, S.~Ransom, M.~Roberts and J.~Hessels,
  Nature {\bf 467}, 1081 (2010).
\bibitem{Antoniadis:2013pzd}
  J.~Antoniadis {\it et al.},
  Science {\bf 340}, 6131 (2013).
\bibitem{Abadie:2010cf}
  J.~Abadie {\it et al.} [LIGO Scientific and VIRGO Collaborations],
  Class.\ Quant.\ Grav.\  {\bf 27}, 173001 (2010).
\bibitem{LIGO:Data} LIGO Open Science Center. Retrived from https://losc.ligo.org/S6

\end{thebibliography}
\end{document}